# Design Patterns as Quality Influencing Factor in Object Oriented Design Approach

[1]Poornima. U. S., [2] Suma. V, [3]Vasanth Kumar. H
[1,3]RR Institute of Technology,
[1,2]Research and Industry Incubation Centre,
Dayananda Sagar Institutions, Bangalore
[1]uspaims@gmail.com, [2]sumavdsce@gmail.com, vasekumar@gmail.com

*Abstract*— Object Oriented Design methodology is an emerging software development approach for complex systems with huge set of requirements. Unlike procedural approach, it captures the requirements as a set of data rather than services, encapsulated as a single entity. The success of such a project relies on major factors like design patterns, framework, key principles, metric standards and best practices adapted by the industry. The patterns are key structures for a recursive problem bits in the problem domain. The combination of design patterns forms a framework which suits the problem statement in hand. The pattern includes static design and dynamic behavior of different types of entities which can be mapped as a functional diagram with cardinalities between them. The degree of cardinality represents the coupling factor which the industry perceives and measures for software design quality. The organization-specific design principles and rich repository of on-the-shelf patterns are the major design-quality-influencing-factors contribute to software success. These are the asset of an industry to deliver a quality product to sustain itself in the competitive market.

## I. INTRODUCTION

In recent years, Industry is receiving a verity of problem statements from different domains prone to enhance in the nearest future. The project success rate depends on the development process and quality standards adapted by the industry. Apart from other phases, design phase is a crucial one which demands more attention from the development team, contributing to the project success and enhancement in the future. However, the error rate in design phase is increasing along with the system complexity [1]. To reduce the flaws at design phase, the industries are tackling on the design problems by working with the existing design practices, models and patterns rather than a new approach. Object-Oriented Development Methodology is a popular project development approach for complex and data-centric problems conceive the solution in a realistic way.

Patterns are reusable design components in software industry which are framed based on the experience gained across different projects. Periodical refinements of the patterns are done to keep the best patterns on the shelf as an asset of the industry. Patterns are taking quick reformation since industries are handling variety of problem statements. However, design pattern quality is based on architecture team cognitive power and past experience with same kind of projects they dealt with.

Patterns are one of the major design influencing factors to account the quality of an end product. It presents static and dynamic nature of the solution model of a problem domain. In Object Orientated (OO) systems, classes, objects, interfaces, delegates, properties represent the elements of the solution space and pattern constitute interrelationship among them to make the design more flexible and reusable [2][3].

## II. OBJECT ORIENTED DESIGN PATTERNS

Work-flow model is a primary step towards software development in an industry. It is a conceptual pattern of overall system which provides different components of a project under development. Patterns are related to a context depicting a component structure which renders the static and dynamic relationship between the components in an application space. Such designs are implemented using suitable programming constructs/patterns at the later stage of software development.

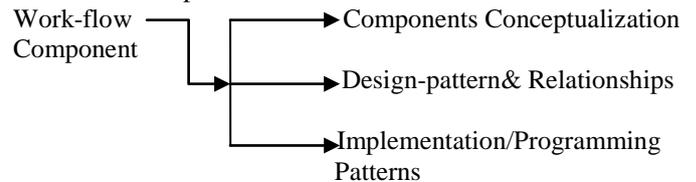

Fig.1. Pattern of software development process.

It is a recursive process for each workflow components in an application domain to build a model, further the design patterns and relationships plays an important role in success of a project [4].

The pattern for a component is perceived by major activities called 'three-rules for pattern design' which systematically constructs a pattern for a context in a problem domain.

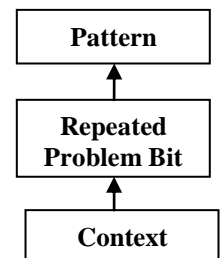

Fig.2. Bottom-Up approach for pattern inception.

Contexts are the output of the workflow process during the analysis of a system. Each context yields with problem bits which are repeatedly occurring in the software development. A proven resolution for such bits are designed as patterns and kept as an asset in the repository in an industry for future reference.

### III. STATIC AND DYNAMIC DESIGN PATTERNS:

Design patterns are creational, structural and behavioral. Creational and structural represents the static design and behavioral the dynamic facet of the design. The static pattern focuses on the architectural elements like class, interface, enumerations and structures related to a programming language. However the dynamic pattern relates the elements to show the behavior of the overall system.

There are more than 23 patterns popular and in practice in industry. They provide an insight to an architect in designing the system [5][6][7][8].

| Creational Design Pattern | Structural Design Pattern | Behavioral Design Pattern |
|---|---|---|
| Abstract Factory | Adapter | Chain of Responsibility |
| Builder | Bridge | Observer, Command |
| Factory Method | Composite | State, Strategy |
| Prototype | Decorator | Iterator Interpreter |
| Singleton | Façade | Mediator, Visitor |
|  | Flyweight | Template Method |
|  | Proxy | Memento |

Table.1. Static and Dynamic patterns.

#### A. Creational pattern: A static design approach

Creational patterns support to build the architecture of a bit of problem statement. It is framed from simple to complex bit and the success depends on the right usage of patterns.

*a. Singleton: A pattern with a class instantiated to single object:*

A simple pattern with single object which provides global access in a project. It is applicable in a situation where single common functionality like configuration, login to a remote system are done through a common accessible object.

*b. Factory: Client gets object without knowing implementation*

It is very popular pattern where client request a product by supplying the required features to a factory class. It creates and sends it back to the caller by hiding the implementation details. Such classes are easy to upgrade as it is not affecting the client object.

*c. Factory Method Pattern: Object type is left to the client:*

In this pattern, an interface provides a common platform to for all the subclasses and it is left to them to pass the class type to instantiate during run time.

*d. Abstract Factory: Interface for creating related objects:*

It provides interfaces for creating a family of related objects without specifying their classes explicitly. It is helpful when system is configured to work with multiple families of products. Single request is enough to initiate set of actions to serve the purpose.

*e. Builder: Creates complex object:*

The complexity of classes and objects are proportional to the requirements density in the solution space. An object is more complex in nature as it is a part of containment. Builder is an interface creates such objects separate from the class depending on the information given by the subclasses. Thus it supports 'separation of concern' reducing the load of the super class.

*f. Prototype: Clone the object:*

The projects need to be cost effective and required to deliver high performance. Prototype pattern supports the same by creating a clone object whenever new object is required. When client requires an object, it sends a request to prototype class which in turn calls clone () method to create an object.

#### B. Structural patterns:

This pattern category makes changes in to the classes so that they can be better utilized in implantation. The adapter pattern converts the interface of one class to another which client expects, where as flyweight pattern creates an interface shared by objects whose internal state is common. Decorator pattern increases the functionality at run time and Bridge pattern separates the abstraction from the implementation so that both can vary independently.

#### C. Behavioral Pattern: A dynamic design approach:

These patterns show the dynamic facet of the system. The objects interaction between the class represents the system behavior and need to be optimized for better design quality.

*a. Chain of responsibility: Series of action execution:*

To handle even generated, objects need to interact each other with a command without knowing the participation each other. The related object will respond in a chain to complete the even initiated by the client.

*b. Command: Interface to execute commands:*

In this pattern, an interface for executing the commands is provided to the clients. When client requests an operation, it is encapsulated and forwarded to the command class for execution. The method execute() invokes the correct operation with the help of related methods belongs to both server and client.

*c. Iterator: List of objects as in arry:*

This pattern resembles the concept of array data structure where elements are executed sequentially. This pattern provides an interface where objects of similar kind can be executed sequentially more than once as requested by the client.

*d. Mediator : Loose coupling:*

A good object Oriented design provides a collect of objects interacting each other without aware of their complete details. Thus the system provides loose coupling and scalability of the system is increased.

*e. Observer : Dependency of objects:*

This is an important pattern defines the dependency between the objects in a scenario. When an object changes the state, it is notified to other dependent objects in the same scenario.

*f. Strategy: Encapsulates algorithms:*

Some classes are differing only in services/algorithms. This pattern defines a class with all algorithms which are encapsulated so that client can request the necessary one when needed.

## IV. DESIGN PATTERN TEMPLTE AND DOCUMENTATION.

As industry is dealing with different problem statements, the pattern of the solution model is also changing to provide the best reusable design. Though different patterns are in practice, there is common decision on documenting the pattern design template so that it could be better modified in future to suit the current scenario. The documentation template being used in industry is similar kind which explores the complete details on the pattern being in practice [9][10][11][12][13].

| Term | Description |
|---|---|
| Pattern Name | Explains the pattern in short with suitable name. |
| Intent | Describes what the pattern does |
| Motivation | Explanation with suitable example |
| Applicability | Lists the situation where the pattern is applicable |
| Structure | Set of classes and objects that depicts the pattern |
| Participants | Set of classes and objects participate in the scenario |
| Collaboration | Describes how participants collaborate to carry out the responsibilities |
| Consequences | Describes the forces exist with the pattern , benefits and trade-offs. |

Table.2. Documentation of pattern design.

## V. DESIGN PATTERNS AND FRAMEWORKS

Software design patterns are templates for commonly occurring problem in a software design. It describes the form of Classes, Objects, Inheritance, Aggregation and the communication between the objects and classes for a particular context. The framework is a collection of such patterns for a problem statement defining the high level abstraction and good communication between designer and user. Several design patterns are in practice and a good pattern is one which makes the overall software design more flexible, elegant and reusable.

The most popular and commonly used pattern is Model View Controller (MVC) which separates the business logic from the user interface. The Model represents the data information present in the application and the business rules used to manipulate the data. The View corresponds to elements of the user interface such as text, checkbox etc., and the Controller manage details involving the communication between the model and view. The controller handles user actions such as keystrokes and mouse movements and pipes them into the model or view as required. Thus both view and controller are dependent on the model, but model is independent of both of them.

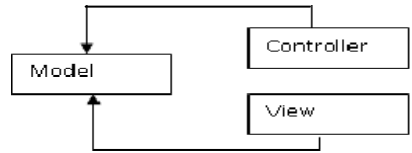

Fig.3. General structure of MVC Model.

However, the model could be implemented in two ways. In passive model, the controller changes the model and informs the view to get refreshed. The model need not inform the other two, regarding the changes happened.

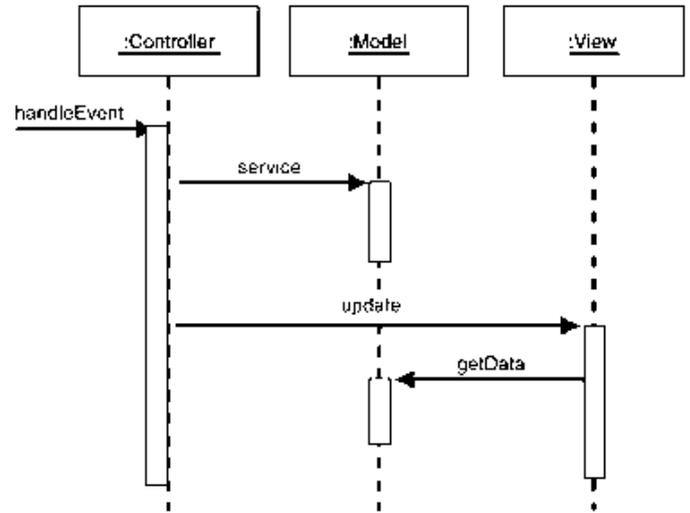

Fig.4. Functionality of passive model.

Active model is used when model changes its state without controller's involvement. When sources are changing data at client side, it must be reflected at views. This separation is achieved by using *Observer pattern* in this model so that objects change the state without defining their dependencies.

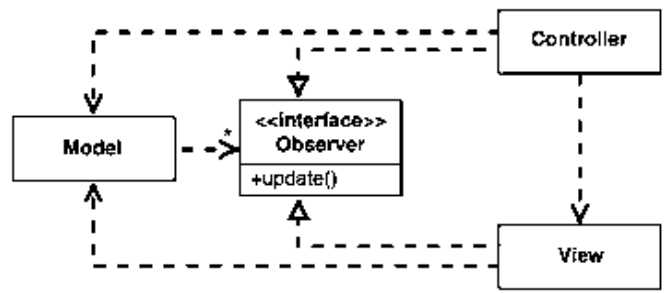

Fig.5. Introducing Observer in active model to separate view from model.

Thus the active model achieves the separation of logical schema from external view there by providing logical data independence.

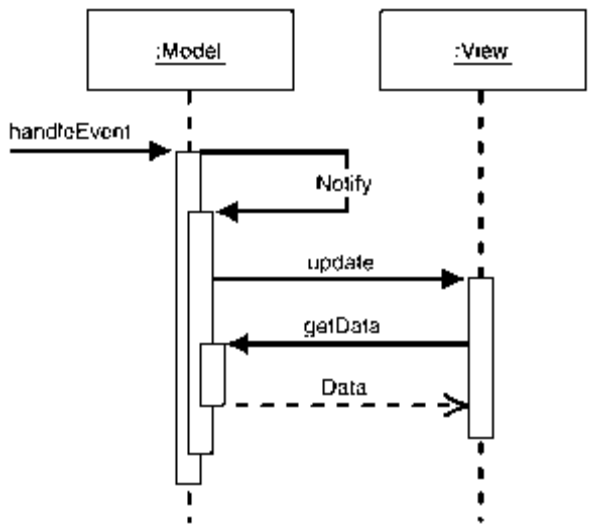

Fig.6. Functionality of active model.

## VI. CONCLUSION

Object Orientation is a popular software development for addressing huge problem domain with verity of requirements. The success of such projects relies on the quality of overall design from modelling till quality assessment metrics. Though various patterns are available, it is the architect's cognitive ability to mix and match the patterns to suit the current problem statement in hand. The overall design quality basically depends on the complexity of the dependency between the modules. The creational and structural patterns presents the static design on of a system where as the behavioral patterns provide the interaction between the objects in the dynamic solution space. The design quality depends on the degree of coupling between the objects involved in the interaction. One of the design principles of the pattern design is to minimize the coupling factor for reducing the design complexity and making the pattern simpler. Thus, the complexity of design also depends on patterns been used in the architecture. Right choice of pattern to create the framework increases the overall design quality there by supporting the quality of end product. The industry is looking for better patterns, architectural models to deliver a quality product to satisfy the clients in order to retain the image in competitive market.